\documentclass[twocolumn]{aastex6}

\shorttitle{Resolving Close Encounters}
\shortauthors{Stephen R. Kane}
\slugcomment{Submitted for publication in the Astrophysical Journal}

\begin{document}

\title{Resolving Close Encounters: Stability in the HD~5319 and
  HD~7924 Planetary Systems}

\author{Stephen R. Kane}
\affil{Department of Physics \& Astronomy, San Francisco State
  University, 1600 Holloway Avenue, San Francisco, CA 94132, USA}
\email{skane@sfsu.edu}

%%%%%%%%%%%%%%%%%%%%%%%%%%%%%%%%%%%%%%%%%%%%%%%%%%%%%%%%%%%%%%%%%%%%

\begin{abstract}

Radial velocity searches for exoplanets have detected many
multi-planet systems around nearby bright stars. An advantage of this
technique is that it generally samples the orbit outside of
inferior/superior conjunction, potentially allowing the Keplerian
elements of eccentricity and argument of periastron to be well
characterized. The orbital architectures for some of these systems
show signs of close planetary encounters that may render the systems
unstable as described. We provide an in-depth analysis of two such
systems: HD~5319 and HD~7924, for which the scenario of coplanar
orbits results in their rapid destabilization. The poorly constrained
periastron arguments of the outer planets in these systems further
emphasizes the need for detailed investigations. An exhaustive scan of
parameter space via dynamical simulations reveals specific mutual
inclinations between the two outer planets in each system that allow
for stable configurations over long timescales. We compare these
configurations with those presented by mean-motion resonance as
possible stability sources. Finally, we discuss the relevance to
interpretation of multi-planet Keplerian orbits and suggest additional
observations that will help to resolve the system stabilities.

\end{abstract}


%%%%%%%%%%%%%%%%%%%%%%%%%%%%%%%%%%%%%%%%%%%%%%%%%%%%%%%%%%%%%%%%%%%%

\section{Introduction}
\label{intro}

Detection of multi-planet systems via the radial velocity (RV) method
are becoming increasing common as both the duration and sensitivity of
RV surveys increase. A crucial step in examining these multi-planet
systems is the analysis of the orbital stability of the planets over
long timescales \citep{smi09}. Since the RV technique can sample the
orbit during any position of the orbital phase, it is particularly
well suited to providing information on the eccentricity and argument
of periastron for each planet. These Keplerian orbital elements can
result in orbits that imply close encounters between the planets in
the system. Mean-motion orbital resonances (MMR), such as the Pluto
2:3 resonance with Neptune, can prevent close encounters and result in
orbital stability (for example, see \citet{bar06a,bar07} and
references therein). Dynamical simulations of planetary systems with
an additional planet inserted at an arbitrary semi-major axis are
often used to numerically determine the location of the islands of
stability at the mean-motion resonances \citep{kan15}.

Two planetary systems were recently moved from single-planet to
multi-planet status through additional observations. The star HD~5319
was found to harbor a planet by \citet{rob07}, and then an additional
planet by \citet{gig15}. The first planet in the HD~7924 was
discovered by \citet{how09}, after which the system was expanded with
two more planets by \citet{ful15}. Based upon the published orbital
parameters, both of these systems exhibit evidence of orbital
instability due to potential close encounters of the two outer
planets. The inclination of the planetary orbits to the plane of the
sky for these systems is unknown, and so a dynamical solution to
avoiding close encounters may include mutual inclinations between the
orbits in addition to MMRs that may be present.

In this paper, we present dynamical simulations of the HD~5319 and
HD~7924 systems that help to resolve possible close encounters of the
outer planets via mutual inclinations of the planetary
orbits. Section~\ref{system} presents a description of the problem
being addressed, including the relevant system parameters and
quantifying the proximity of the planetary orbits to each
other. Section~\ref{method} outlines the methodological approach and
parameters used in the numerical simulations. The results for the
HD~5319 and HD~7924 simulations are presented in Sections \ref{hd5319}
and \ref{hd7924} respectively. Section~\ref{mmr} investigates the
possibility of mean-motion resonances as additional sources of
stability. Section~\ref{obs} investigates the effect of periastron
argument on system stability and details a strategy for further
observations that could help to resolve the orbits of the planetary
systems. We provide concluding remarks in Section~\ref{conclusions}.

%%%%%%%%%%%%%%%%%%%%%%%%%%%%%%%%%%%%%%%%%%%%%%%%%%%%%%%%%%%%%%%%%%%%

\begin{figure*}
  \begin{center}
    \begin{tabular}{cc}
      \includegraphics[angle=270,width=8.5cm]{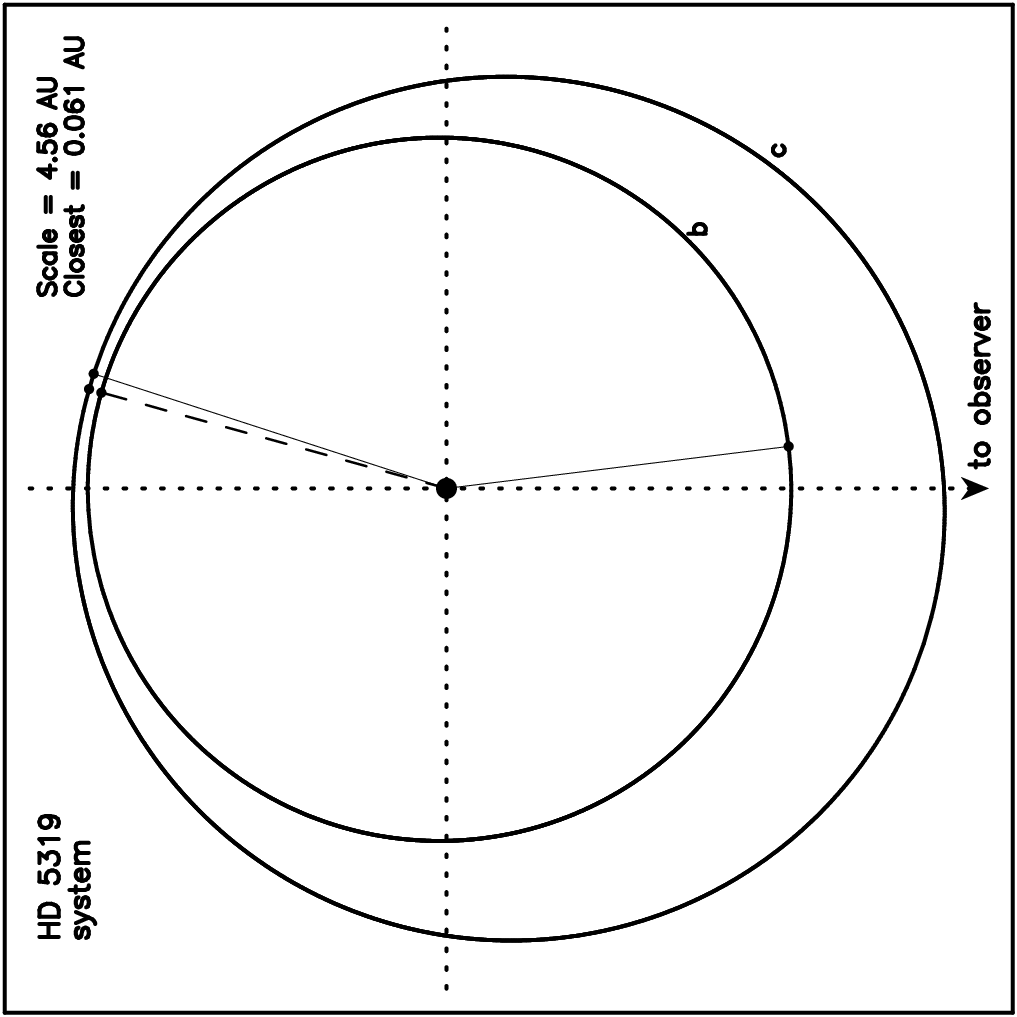} &
      \includegraphics[angle=270,width=8.5cm]{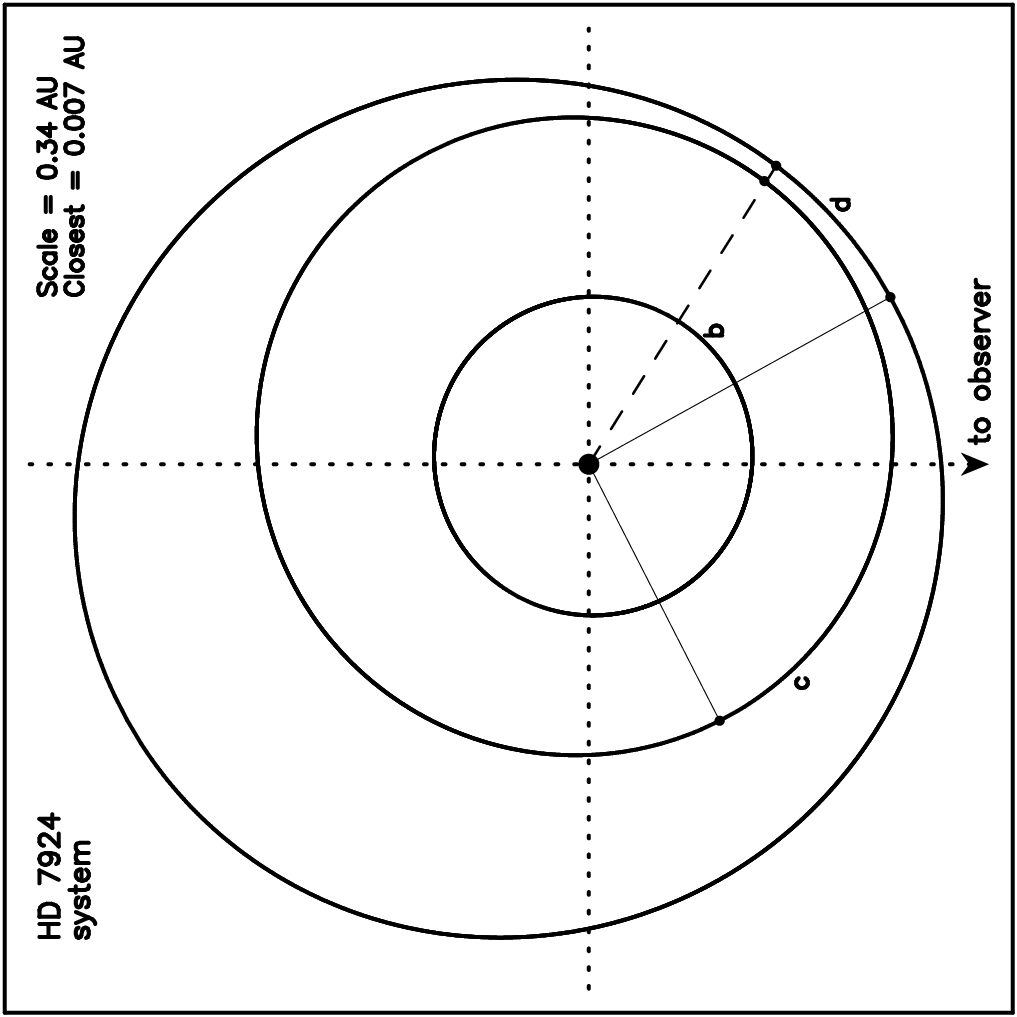}
    \end{tabular}
  \end{center}
  \caption{Top-down views of the HD~5319 (left) and HD~7924 (right)
    planetary systems. The orbits have been plotted using the system
    parameters shown in Table~\ref{planets}. The ``scale'' refers to
    the scale of the plot along a side, and ``closest'' refers to the
    closest approach of the two outer planet orbits assuming they are
    coplanar. The solid lines joining host star to the orbits shows
    the periastron location for each orbit. The dashed line joining
    host star to the orbits indicates the location of closest
    approach.}
  \label{orbits}
\end{figure*}

\section{System Parameters}
\label{system}

The orbital and physical characteristics of the HD~5319 and HD~7924
planets were extracted from the publications by \citet{gig15} and
\citet{ful15} respectively. In the case of HD~5319, \citet{gig15} used
the Keplerian Fitting Made Easy package \citep{gig12} to produce the
final orbital solution, then validated the solution with a
Differential-Evolution Markov Chain Monte Carlo (DE-MCMC) approach
that included a 100 year dynamical stability constraint. \citet{gig15}
also performed stability simulations that showed the majority of the
DE-MCMC results were unstable over $10^7$ years with the exception of
a subset of the results at the 4:3 MMR. For HD~7924, \citet{ful15}
used the RVLIN package \citep{wri09} to estimate the initial orbital
solution and then used DE-MCMC to produce the final solution. The
median fit parameters from the DE-MCMC analysis were used for a single
stability simulation that proved to be stable for $10^5$ years. The
planetary parameters derived from these methods that are relevant to
our analysis are shown in Table~\ref{planets}, including the mass of
the host star ($M_\star$), orbital period ($P$), time of periastron
passage ($T_p$), orbital eccentricity ($e$), argument of periastron
($\omega$), and semi-major axis ($a$). The orbits of the planets in
the HD~5319 and HD~7924 systems are depicted in the left and right
panels of Figure~\ref{orbits} respectively. Also shown in the panels
are solid lines from the host star that represent the periastron
arguments for the orbits.

\begin{deluxetable*}{lccccc}
  \tablecolumns{7}
  \tablewidth{0pc}
  \tablecaption{\label{planets} Stellar and Planetary Parameters}
  \tablehead{
    \colhead{Parameter} &
    \multicolumn{2}{c}{HD~5319 ($M_\star=1.51 \pm 0.11 \ M_\odot$)} &
    \multicolumn{3}{c}{HD~7924 ($M_\star=0.832^{+0.022}_{-0.036} \ M_\odot$)} \\
    \colhead{} &
    \colhead{b} &
    \colhead{c} &
    \colhead{b} &
    \colhead{c} &
    \colhead{d}
  }
  \startdata
  $P$ (days)     & $641 \pm 2$           & $886 \pm 8$ &
  $5.39792\pm0.00025$ & $15.299^{+0.0032}_{-0.0033}$ &
  $24.451^{+0.015}_{-0.017}$ \\
  $T_p\,^{1}$  & $6288 \pm 720$        & $3453 \pm 92$ &
  $5586.38^{+0.086}_{-0.110}$ & $5586.29^{+0.40}_{-0.47}$ &
  $5579.1^{+1.0}_{-0.9}$ \\
  $e$            & $0.02 \pm 0.03$       & $0.15 \pm 0.06$ &
  $0.058^{+0.056}_{-0.040}$ & $0.098^{+0.096}_{-0.069}$ &
  $0.21^{+0.13}_{-0.12}$ \\
  $\omega$ (deg) & $97 \pm 90$           & $252 \pm 34$ &
  $332^{+71}_{-50}$ & $27^{+52}_{-60}$ & $119^{+210}_{-97}$ \\
  $M_p \sin i$   & $1.76 \pm 0.07 \ M_J$ & $1.15 \pm 0.08 \ M_J$ &
  $8.68^{+0.52}_{-0.51} \ M_\oplus$ & $7.86^{+0.73}_{-0.71} \ M_\oplus$ &
  $6.44^{+0.79}_{-0.78} \ M_\oplus$ \\
  $a$ (AU)       & $1.6697 \pm 0.0036$   & $2.071 \pm 0.013$ &
  $0.05664^{+0.00067}_{-0.00069}$ & $0.1134^{+0.0013}_{-0.0014}$ &
  $0.1551^{+0.0018}_{-0.0019}$ \\
  $R_H$ (AU) & 0.1202 & 0.1292 & 0.0012 & 0.0024 & 0.0031
  \enddata
  \tablenotetext{1}{JD -- 2,450,000}
\end{deluxetable*}

An additional quantity we calculated for each planet was the Hill
radius, given by
\begin{equation}
  R_H = r \left( \frac{M_p}{3 M_\star} \right)^{1/3}
  \label{hillradius}
\end{equation}
where $r$ is the time-dependent (for a non-circular orbit)
star--planet separation. This separation is given by
\begin{equation}
  r = \frac{a (1 - e^2)}{1 + e \cos f}
  \label{separation}
\end{equation}
where $f$ is the true anomaly. We include the mean Hill radius (where
$r = a$) for each planet in Table~\ref{planets}.

We calculated star--planet separations through the entire orbit for
each planet and determined the location of closest proximity for the
outer planet orbits, assuming that the orbits are coplanar. This
location is indicated by a dashed line in each of the panels of
Figure~\ref{orbits}. For the HD~5319 system, the closest proximity of
the planetary orbits occurs at a star--planet separation and true
anomaly of $r = 1.701$~AU and $f = 158.3\degr$ for planet b, and $r =
1.761$~AU and $f = 3.3\degr$ for planet c. The separation of the
orbits at this location is 0.061~AU, equivalent to 0.495 $R_H$ for
planet b and 0.571 $R_H$ for planet c, where the Hill radii were
calculated using Equation~\ref{hillradius} at the location of closest
approach. Similarly for the HD~7924 system, occurs at $r = 0.118$~AU
and $f = 120.4\degr$ for planet c, and $r = 0.125$~AU and $f =
28.2\degr$ for planet d. In this case the minimum separation between
the outer planet orbits is 0.007~AU, corresponding to 2.749 $R_H$ for
planet c and 2.933 $R_H$ for planet d. Note that the HD~7924 system is
smaller scale than the HD~5319 system, both in terms of planetary
masses and semi-major axes. The proximity of the orbits in each case
emphasizes the need for detailed dynamical simulations to resolve
potential close planetary encounters.

%%%%%%%%%%%%%%%%%%%%%%%%%%%%%%%%%%%%%%%%%%%%%%%%%%%%%%%%%%%%%%%%%%%%

\section{Methodology}
\label{method}

To assess orbital stability of the planetary systems, we made use of
the Mercury Integrator Package, as described by \citet{cha99}. The
code performs N-body integrations based upon user-specified input
parameters and starting conditions for the system. Our dynamical
simulations use the hybrid symplectic/Bulirsch-Stoer integrator with a
Jacobi coordinate system, which generally provides more accurate
results for multi-planet systems \citep{wis91,wis06}.

For each system, we set up initial conditions using the parameters
shown in Table~\ref{planets}. Each of the integrations were performed
for a simulation duration of $10^7$ years commencing at the present
epoch. The time resolution for the simulations were chosen to meet the
minimum timestep recommendation of \citet{dun98}: $1/20$ of the
shortest orbital period in the system. To meet this requirement, we
used timesteps of 5 days and 0.25 days for the HD~5319 and HD~7924
systems respectively. Results from each integration were output in
steps of 100 years.

We conducted a single simulation for each system assuming that the
orbits are approximately coplanar ($i = 90\degr$), verifying that the
systems are indeed unstable as described. We then extended our
analysis by running an exhaustive set of simulations that slowly
changed the orbital inclination of the outer planet, from $i =
90\degr$ to $i = 60\degr$ in steps of $0.1\degr$. This introduces a
mutual inclination between the two outer planets that allows a search
for islands of stability that may resolve the close encounter dilemma.

The orbital parameters described in Section~\ref{system} have
associated uncertainties that may also account for system
stability. To investigate this, we conducted additional simulations
that vary the argument of periastron to locate islands of stability
for the coplanar scenario. These results are presented in the context
of refining the orbital parameters in Section~\ref{obs}.

%%%%%%%%%%%%%%%%%%%%%%%%%%%%%%%%%%%%%%%%%%%%%%%%%%%%%%%%%%%%%%%%%%%%

\section{The HD~5319 System}
\label{hd5319}

\begin{figure*}
  \begin{center}
    \includegraphics[angle=270,width=15.0cm]{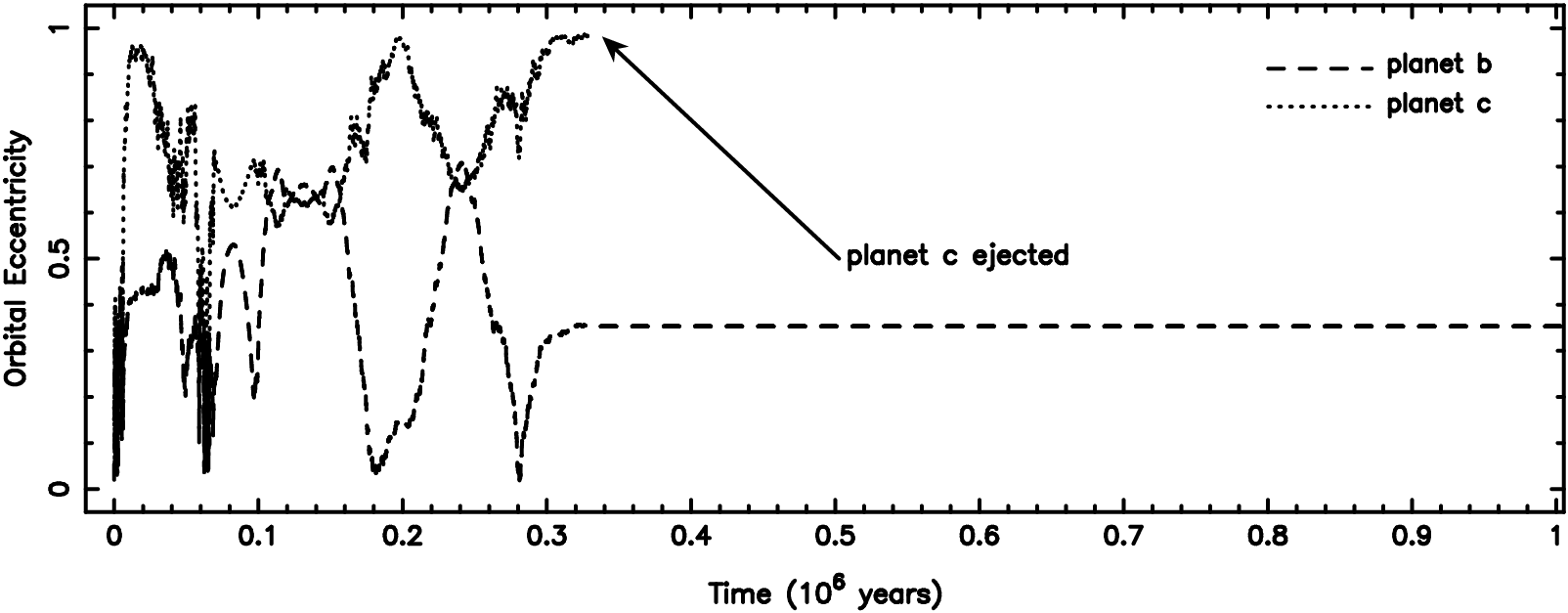} \\
    \hspace{0.5cm}\includegraphics[angle=270,width=16.0cm]{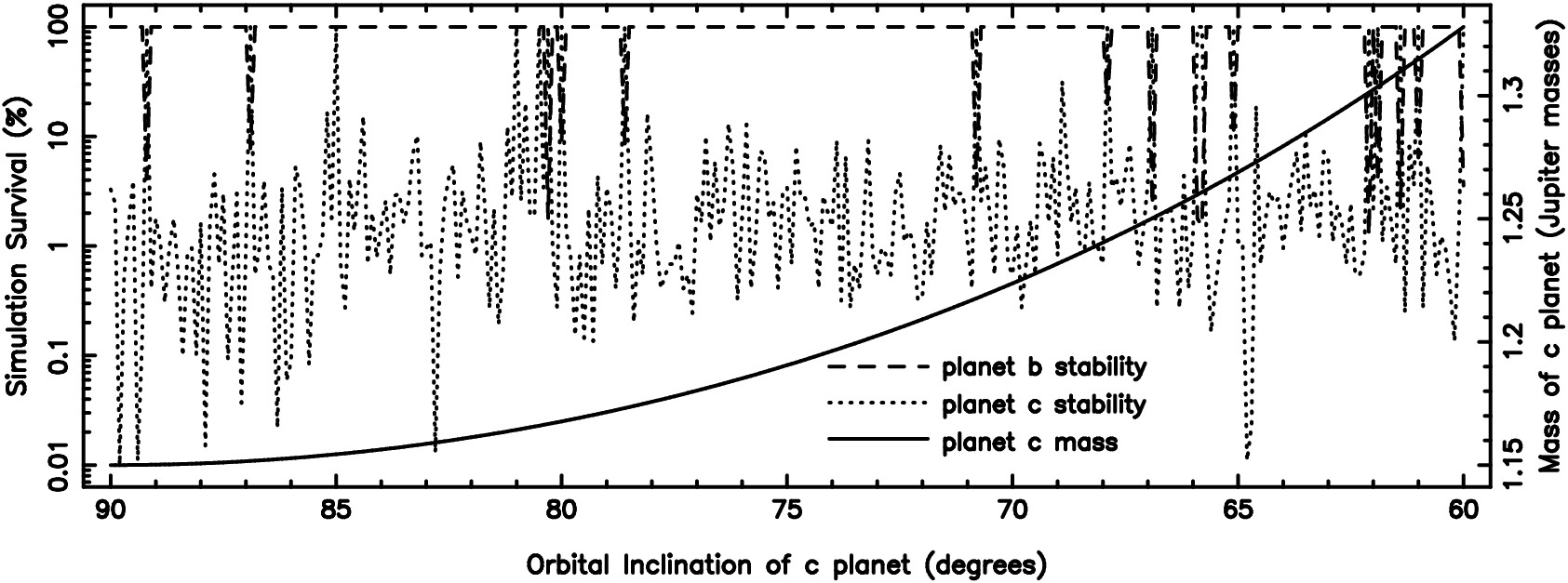} \\
    \includegraphics[angle=270,width=15.0cm]{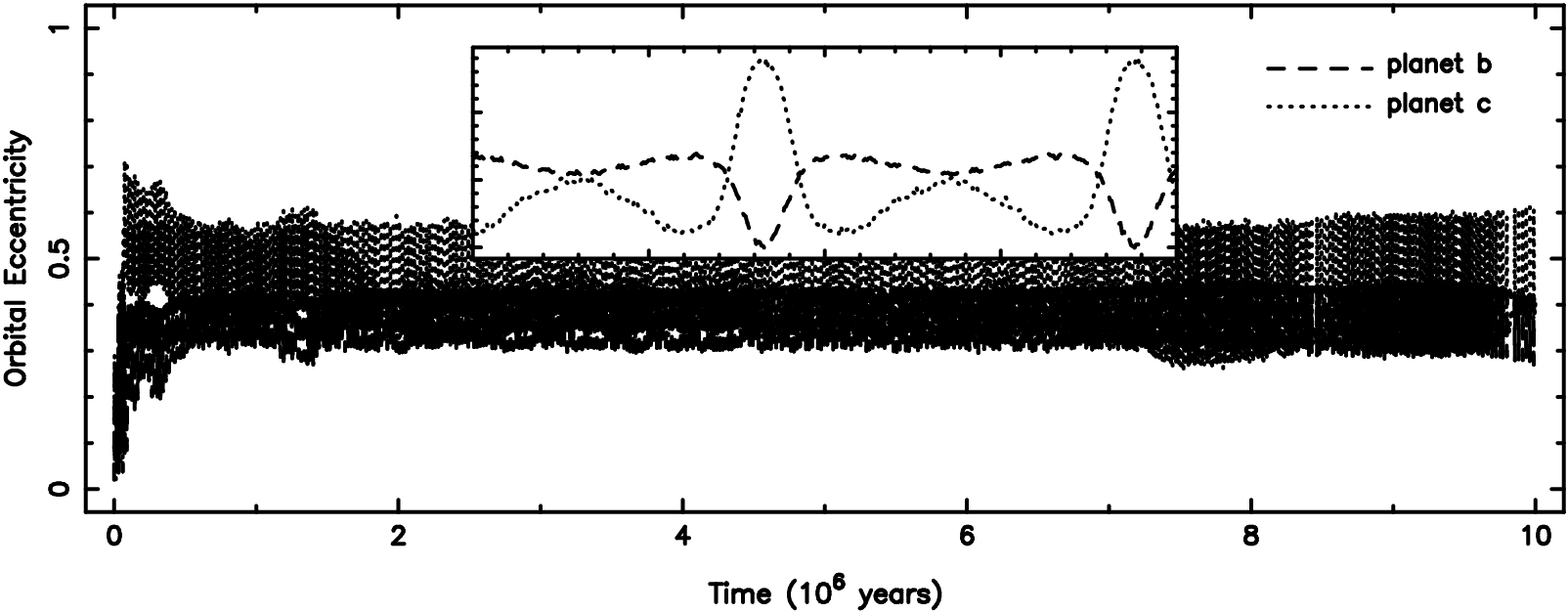}
  \end{center}
  \caption{Top: The orbital stability of the HD~5319 system when the
    orbits are assumed to be coplanar at $i = 90\degr$, represented by
    the evolution of the orbital eccentricities. Planet c is ejected
    after 328,500 years and planet b remains with an eccentricity of
    $e = 0.35$. Middle: The percentage simulation survival of the
    HD~5319 planets as a function of orbital inclination of planet
    c. The solid line represents the increasing true mass of planet c
    (shown on the right-hand axis) as the orbital inclination
    decreases. Inclinations for which both planets remain in the
    system over the duration of the simulation are considered
    stable. Bottom: Orbital eccentricities as for the top panel, but
    for the case where planet c has an inclination of $i_c =
    85\degr$. The subpanel shows a zoomed 40,000 year segment of the
    eccentricity oscillations that occur during the simulation. Both
    planets remain in stable orbits for the entire $10^7$ year
    simulation duration.}
  \label{hd5319fig}
\end{figure*}

The two planets of the HD~5319 system have their closest approach
where $\omega + f \sim 255\degr$ (see Section~\ref{system}). Our
dynamical stability simulation for the assumption of coplanar orbits
shows that the planetary system can only survive for $\sim$330,000
years based on the system parameters from Section~\ref{system}. The
results of this simulation are shown in the top panel of
Figure~\ref{hd5319fig}, where the dashed and dotted lines represent
the orbital eccentricities of planets b and c respectively. Planet c
remains in the system with an eccentricity of $e \sim 0.35$ after
planet b is ejected.

The results of the simulation that add a mutual inclination between
planets b and c (see details in Section~\ref{method}) are shown in the
middle panel of Figure~\ref{hd5319fig}. The solid line indicates the
changing mass of planet c as the inclination is gradually decreased
from $i_c = 90\degr$. Even with the increased mass of planet c, planet
b remains the dominant mass during close encounters and thus remains
in the system during the majority of inclination cases. We located
three inclinations in the range $90\degr > i_c > 60\degr$ for which
both planets survived the complete simulation duration of $10^7$
years: $85\degr$, $81\degr$, and $80.5\degr$.

The lower panel of Figure~\ref{hd5319fig} shows the variation in
orbital eccentricities for both planets where planet c has an
inclination of $i_c = 85\degr$. The system is able to acquire a stable
configuration whereby angular momentum is transferred between the two
planets, oscillating the eccentricities over long timescales, as
described by \citet{kan14a}. The zoomed inset in the lower panel of
Figure~\ref{hd5319fig} shows the eccentricity oscillations over a
40,000 year segment of the simulation.

%%%%%%%%%%%%%%%%%%%%%%%%%%%%%%%%%%%%%%%%%%%%%%%%%%%%%%%%%%%%%%%%%%%%

\section{The HD~7924 System}
\label{hd7924}

\begin{figure*}
  \begin{center}
    \includegraphics[angle=270,width=15.0cm]{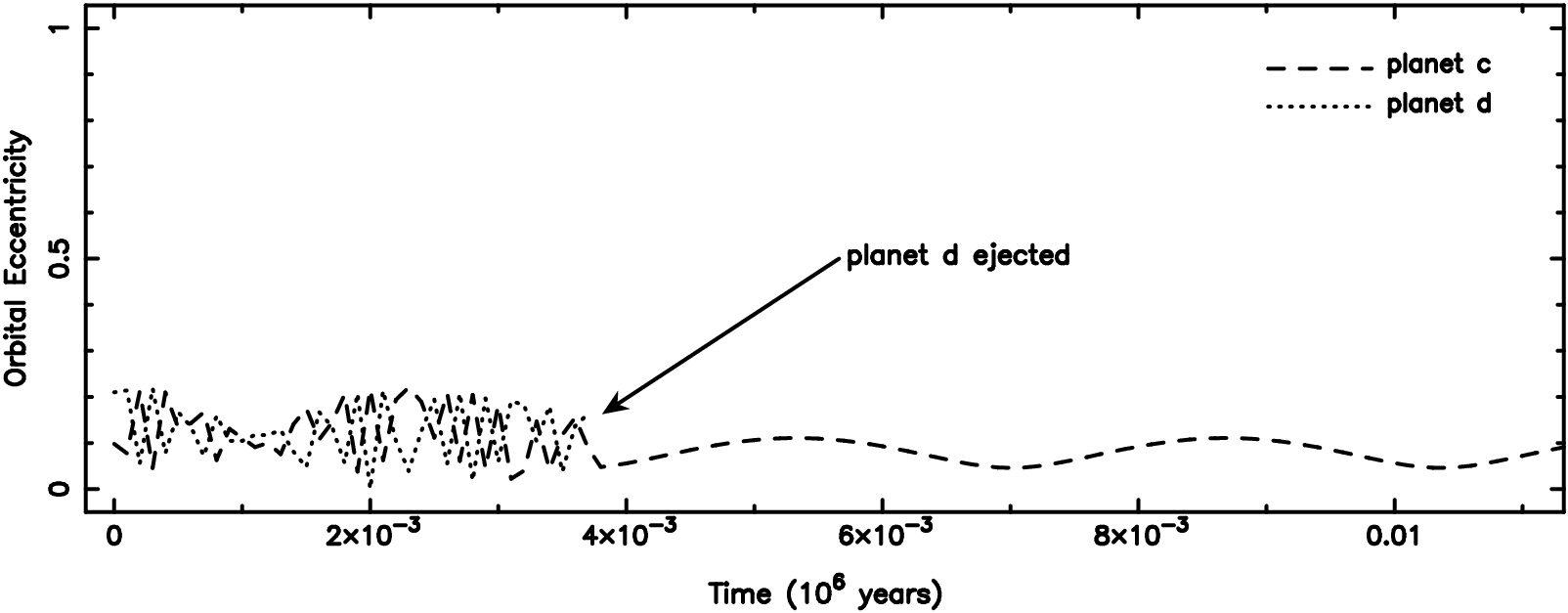} \\
    \hspace{0.5cm}\includegraphics[angle=270,width=16.0cm]{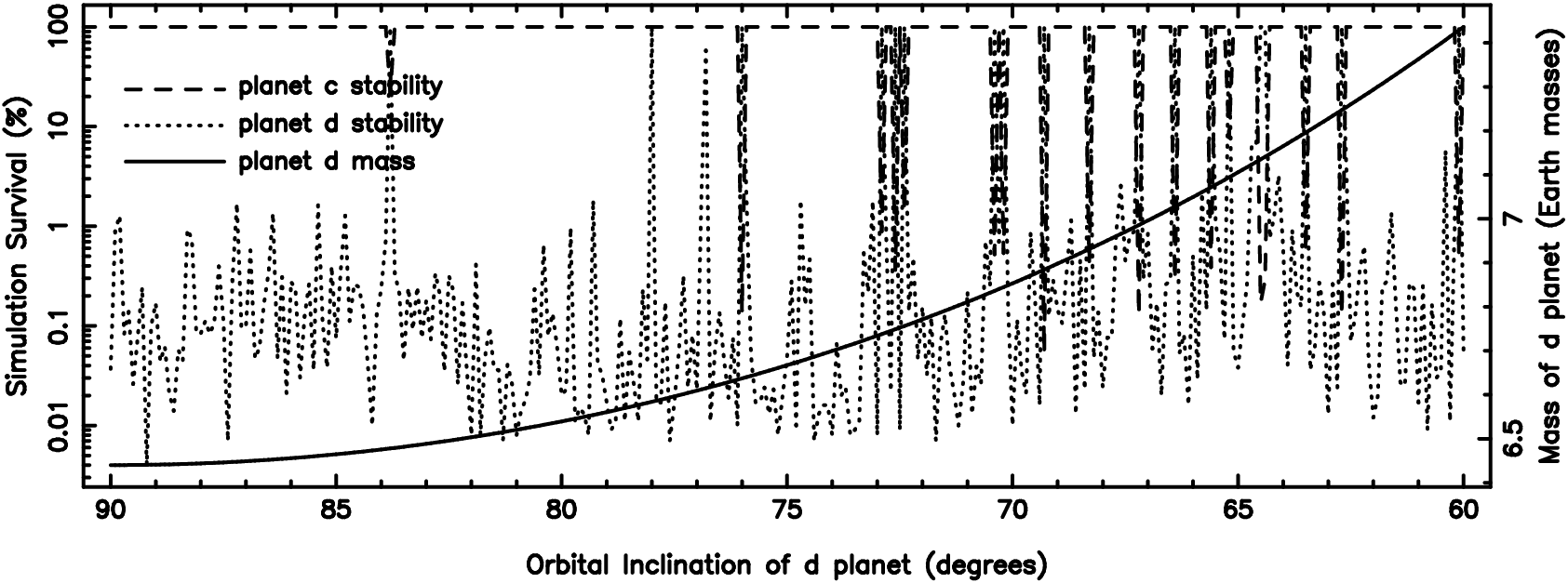} \\
    \includegraphics[angle=270,width=15.0cm]{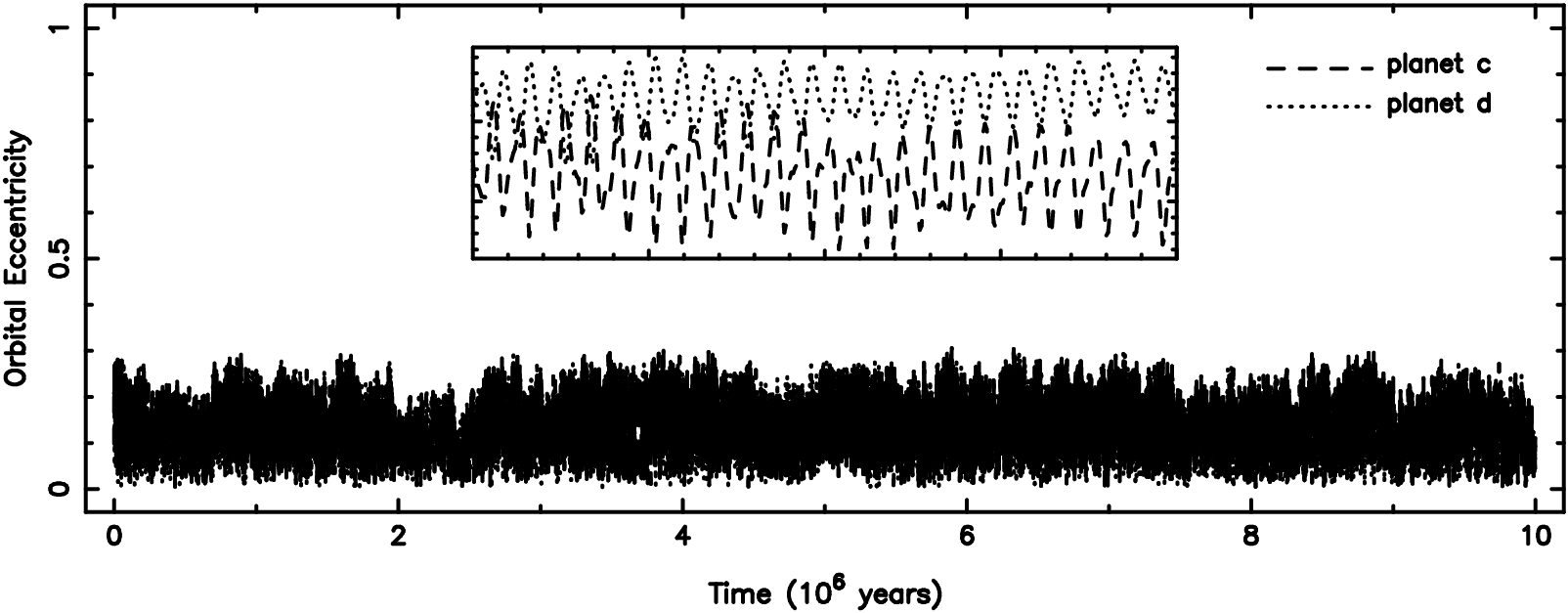}
  \end{center}
  \caption{Top: The orbital stability of the HD~7924 system when the
    orbits are assumed to be coplanar at $i = 90\degr$, represented by
    the evolution of the orbital eccentricities. Planet d is ejected
    after only 3,700 years and planet c remains with an eccentricity
    that oscillates with a mean value of $e = 0.1$. Middle: The
    percentage simulation survival of the HD~7924 planets as a
    function of orbital inclination of planet d. The solid line
    represents the increasing true mass of planet d (shown on the
    right-hand axis) as the orbital inclination
    decreases. Inclinations for which both planets remain in the
    system over the duration of the simulation are considered
    stable. Bottom: Orbital eccentricities as for the top panel, but
    for the case where planet d has an inclination of $i_d =
    78\degr$. The subpanel shows a zoomed 40,000 year segment of the
    eccentricity oscillations that occur during the simulation. Both
    planets remain in stable orbits for the entire $10^7$ year
    simulation duration.}
  \label{hd7924fig}
\end{figure*}

The orbits of the two outer planets (planet c and d) for the HD~7924
system have their closest approach where $\omega + f \sim 147\degr$
(see Section~\ref{system}). The planets of this system are
significantly less massive than those of the HD~5319 system and are
several Hill radii apart at the closest approach. Even so, the
dynamical stability of the system for the coplanar scenario is
disrupted relatively early and planet d is lost after only 3,700
years, as shown in the top panel of Figure~\ref{hd7924fig}. After this
event, planet c oscillates in orbital eccentricity as it exchanges
angular momentum with planet b. Note that, even though \citet{ful15}
found the system to be stable for $10^5$ years, the stability duration
is sensitive to the timestep used. As described in
Section~\ref{method}, we use a timestep of 0.25 days but there are a
small fraction of timesteps in the range 0.05--0.5 days that produce
stable outcomes that last slightly more than $10^5$ years for the same
initial conditions.

Similar to the procedure described in Section~\ref{hd5319}, we
performed simulations that vary the inclination of the outer planet in
the range $90\degr > i_d > 60\degr$. The results of these simulations
are shown in the middle panel of Figure~\ref{hd7924fig}. Within the
searched inclination range, our simulations revealed only one stable
configuration, located at an inclination of $i_d = 78\degr$ for planet
d. The variation in orbital eccentricity for planets c and d are shown
in the bottom panel of Figure~\ref{hd7924fig}, with the zoomed inset
panel showing the angular momentum exchange of the two outer
planets. The two outer planets maintain stability with eccentricities
staying below $\sim$0.3. It should be noted that there are large
uncertainties associated with the periastron argument for planet
d. Thus there could be a more suitable value for $\omega$ that would
allow more stable configurations for the system.

%%%%%%%%%%%%%%%%%%%%%%%%%%%%%%%%%%%%%%%%%%%%%%%%%%%%%%%%%%%%%%%%%%%%

\section{Mean-Motion Resonances}
\label{mmr}

A further consideration are MMRs as potential sources of dynamical
stability. MMRs have been considered in detail by various authors
\citep{var99,pet13,ant14}, including the effect of mutual inclinations
\citep{bar15}. Examples of stable MMRs are 3:2 (1.5), 4:3 (1.33), 5:4
(1.25), 5:3 (1.67), and 8:5 (1.6). From Table~\ref{planets}, the
ratios of the orbital periods for the two outer planets are $1.38 \pm
0.01$ and $1.598 \pm 0.001$ for the HD~5319 and HD~7924 systems
respectively.

\begin{figure*}
  \begin{center}
    \begin{tabular}{cc}
      \includegraphics[angle=270,width=8.5cm]{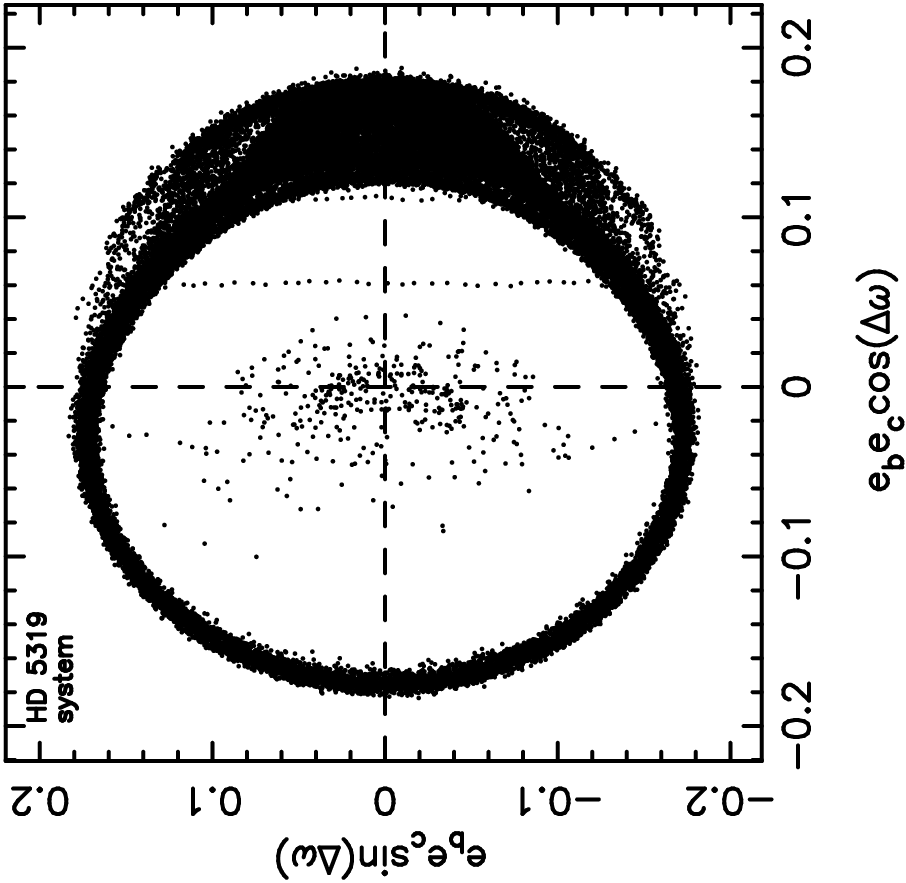} &
      \includegraphics[angle=270,width=8.5cm]{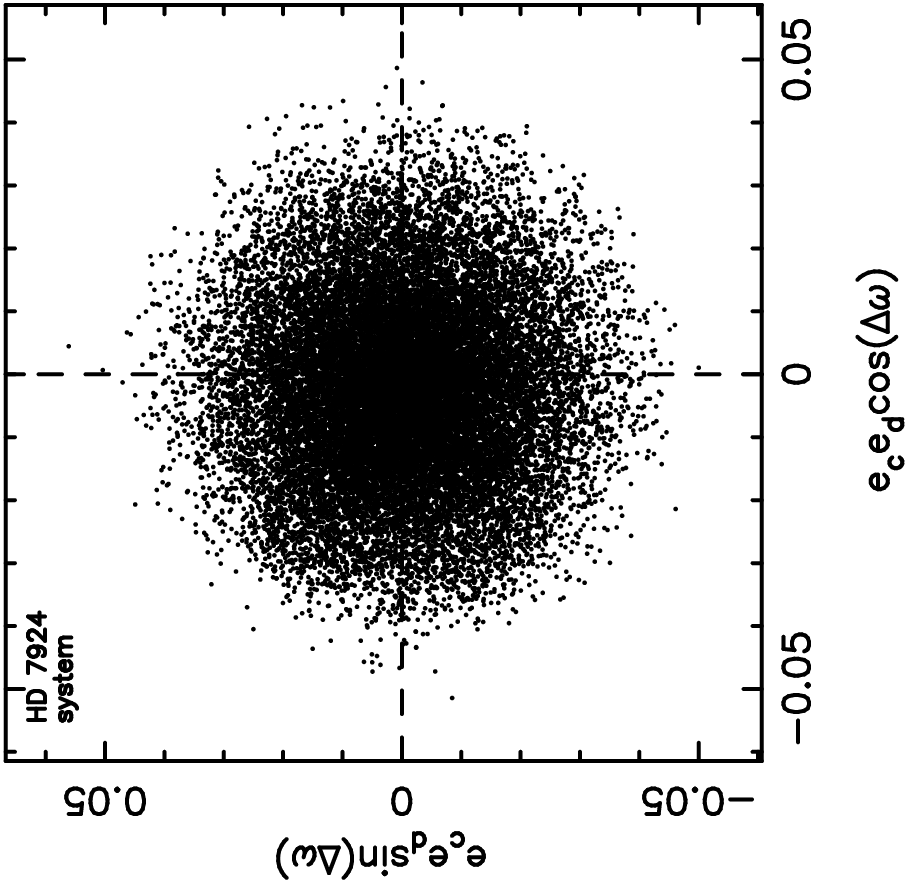}
    \end{tabular}
  \end{center}
  \caption{Apsidal trajectory represented as a polar plot of $e_b e_c$
    versus $\Delta \omega$ for the HD~5319 system (left panel) and
    $e_c e_d$ versus $\Delta \omega$ for the HD~7924 system (right
    panel). These correspond to the stable configurations found in
    Sections~\ref{hd5319} and \ref{hd7924} where $i_c = 85\degr$ for
    HD~5319 and $i_d = 78\degr$ for HD~7924. The apsidal modes for the
    HD~5319 planets move from libration to circulation due to the
    resonance dynamics. For the HD~7924 outer planets, the apsidal
    modes appear to be circulating but include oscillation modes due
    to secular interactions with the inner planet (planet b).}
  \label{epsilon}
\end{figure*}

For the HD~5319 system, the closest MMR is the 4:3 resonance, although
doesnot match to that resonance within the period uncertainties. It
was shown by \citet{bar06b} that the long-term apsidal behavior of
multi-planet orbital elements may be distinguished between libration
and circulation, where the boundary between them is the secular
separatrix. Stability simulations conducted by \citet{gig15} using
coplanar orbits found that several realizations maintained stability
through a mean orbital period ratio close to the 4:3 resonance with a
librating apsidal trajectory. For our stable configuration with $i_c =
85\degr$ (see Section~\ref{hd5319}) we computed the apsidal trajectory
for the HD~5319 system using the eccentricity of the inner and outer
planets ($e_b$ and $e_c$ respectively) and the difference in
periastron arguments ($\Delta \omega$). These are represented
graphically with polar coordinates in the left panel of
Figure~\ref{epsilon}. Clearly the apsidal trajectory evolution of this
system is complex due to both secular and resonant dynamics. The
system begins with librating apsidal modes (the points clustered near
the origin) but moves to circulating apsidal modes since the polar
trajectories encompass the origin. However, the system cannot be
readily classified in terms of libration or circulation.

In the case of the HD~7924 system, the orbital period ratio for the
outer planets is close to the 8:5 MMR. A coplanar stability simulation
by \cite{ful15} was only run for $10^5$ years and did not encounter
stability issues, nor explore resonances. Similarly as for the HD~5319
system, we calculated apsidal trajectories over the $10^7$ year
simulation for the stable configuration with $i_d = 78\degr$. The
resulting polar coordinates using eccentricities of the outer planets
($e_c$ and $e_d$ respectively) are shown in the right panel of
Figure~\ref{epsilon}. The apsidal modes of the system appear to be
circulating although, as noted by \citet{bar06b}, interpretation of
systems with three or more planets is challenging since the secular
interactions of all the planets results in a superposition of the
various oscillation amplitudes. The effect of this is the homogeneous
spreading of data points around the origin, as seen in the right panel
of Figure~\ref{epsilon}.

\begin{figure*}
  \begin{center}
    \includegraphics[angle=270,width=15.0cm]{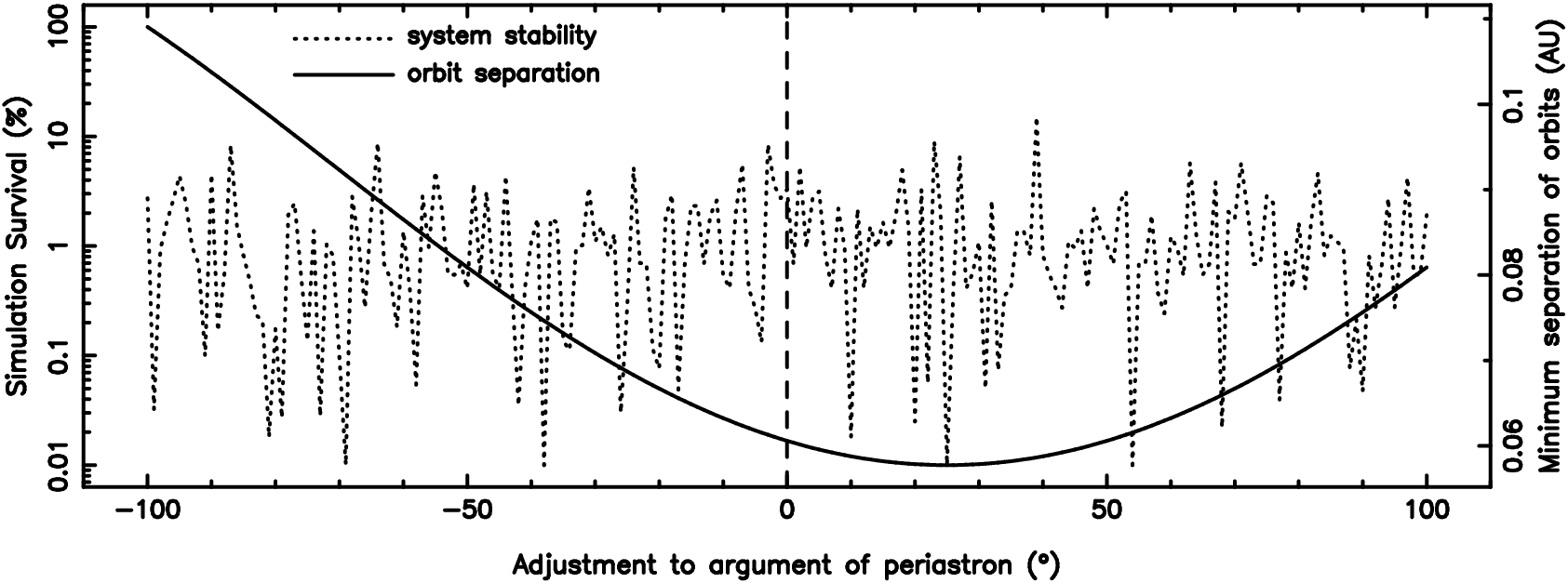} \\
    \includegraphics[angle=270,width=15.0cm]{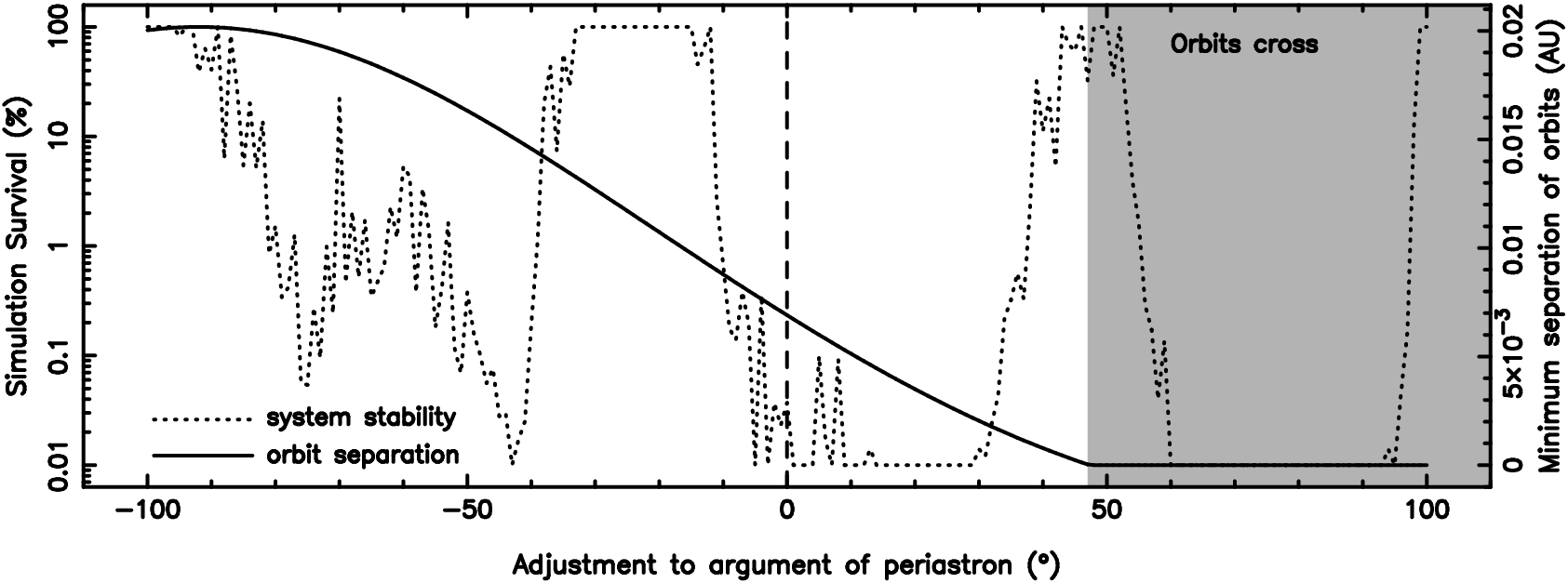}
  \end{center}
  \caption{The stability (dotted line) and minimum separation of the
    outer two planetary orbits (solid line) for the HD~5319 (top
    panel) and HD~7924 (bottom panel) systems. These are represented
    as a function of varying the argument of periastron of the outer
    planet from the measured position, indicated by the vertical
    dashed line. The gray shaded region in the bottom panel represents
    the periastron arguments for which the outer planet orbits cross.}
  \label{vary}
\end{figure*}

An important point to note is that period ratios that diverge slightly
from resonance, such as those described here, do not imply that the
planets concerned do not lie in those resonances. There are in fact
quite few planet pairs that have period ratios occurring exactly at
resonance. It was shown by \citet{gol14} that accounting for ongoing
orbital dynamics explains the distribution of {\it Kepler} planets
found to be near MMR. Additionally, it has been demonstrated by
\citet{gol14} that overstable librations, such as that seen for
HD~5319, can result in eccentricity damping and a subsequent
divergence from the expected value of MMR. The combination of these
factors can lead to an orbital evolution that places the period ratio
preferentially slightly above the expected MMR, as seen in the
dynamical simulations performed by \citet{gig15}.

%%%%%%%%%%%%%%%%%%%%%%%%%%%%%%%%%%%%%%%%%%%%%%%%%%%%%%%%%%%%%%%%%%%%

\section{Further Observations}
\label{obs}

A critical aspect for understanding the stability of a particular
system are the accuracy of the orbital elements. Further RV
observations of known exoplanet systems at calculated optimal times
can provide a dramatic improvement to the planetary orbital parameters
\citep{for08,kan09}. Such improvement is especially necessary for the
eccentricity and argument of periastron; parameters that are often
poorly constrained. For example, the argument of periastron
uncertainties for HD~7924d (see Table~\ref{planets}) are $+210\degr$
and $-97\degr$.

The effect of varying the periastron argument of the outer planet on
the proximity of the outermost planetary orbits are visualized as the
solid line shown in Figure~\ref{vary} for HD~5319 (top) and HD~7924
(bottom). As described in Section~\ref{system}, the minimum separation
of the orbits for the HD~7924 c and d planets is 0.007~AU when using
the measured orbital parameters. Moving the periastron argument in the
negative direction increases the minimum separation of the orbits,
whereas a positive shift decreases the minimum separation. The region
where the periastron argument causes the orbits of the planets to
cross has been shaded gray and the minimum separation fixed at
zero. For mutual inclinations close to zero, such periastron arguments
are usually untenable. The exceptions to this are cases where a MMR can
prevent close encounters even with overlapping orbits, as described by
\citet{mar06}. Since the variation of the argument of periastron can
have such a significant effect on the orbital architecture of the
system, further observations of these systems are highly encouraged to
help resolve the close encounter issue of the outer planets.

To test for system stability as a function of periastron argument, we
conducted additional simulations for each system using the methodology
described in Section~\ref{method}. These simulations assume orbital
coplanarity and vary the periastron argument of the outer planet
between $-100\degr$ and $+100\degr$ in steps of $1\degr$ relative to
the measured values from Table~\ref{planets}, shown as vertical dashed
lines in each panel of Figure~\ref{vary}. Our simulations do not find
any region of stability for the HD~5319 system in the range of
periastron arguments explored for the coplanar case. An additional
parameter to explore would be the variance of eccentricity, and may
possibly explain the stable region found by \citet{gig15} in the
simulation of the same system. For HD~7924, we find several isolated
locations of stability including in the region where the orbits
cross. As described above, such orbital architectures are possible
when MMR is achieved ensuring that the planets do not experience close
encounters over long timescales. The possibility of system stability
at other periastron arguments emphasizes the need for further
observations to refine the orbital parameters of the HD~7924 system.

Planetary multiplicity within a system can be used to constrain
orbital inclination, such as for the HD~10180 system
\citep{lov11,kan14b}. Such constraints are normally determined via
stability simulations that assume coplanarity for the
system. Furthermore, assuming orbital inclinations different from
edge-on (as we have simulated in this work) can result in a modified
dynamical interaction that may be inconsistent with the RV Keplerian
solution. \citet{lau01} and \citet{riv01} investigated this effect
for the GJ~876 system and found that significant divergence with the
RV solution occurs for $\sin i < 0.8$. For our simulations, we explore
the parameter space consisting of $\sin i > 0.866$, with stable
inclinations corresponding to $\sin i = 0.996$ and $\sin i = 0.978$
for HD~5219 and HD~7924 respectively. Thus it is not expected that the
change in masses will cause significant divergence from the best-fit
Keplerian model in our cases.

Ultimately, astrometric data will provide the information required to
understand the true inclinations of the planetary orbits, both with
respect to each other and the plane of the sky. The Gaia mission is an
astrometry mission that was successfully launched by the European
Space Agency (ESA) in 2013. As well as determining a vast number of
stellar parallaxes, the mission will provide astrometry for known
exoplanetary systems in addition to discovering new systems
\citep{per14}. The full capabilities of the Gaia mission are described
in detail by \citet{deb12} and \citet{bai13}.

%%%%%%%%%%%%%%%%%%%%%%%%%%%%%%%%%%%%%%%%%%%%%%%%%%%%%%%%%%%%%%%%%%%%

\section{Conclusions}
\label{conclusions}

RV multi-planet systems provide excellent opportunities to study the
orbital dynamics of Keplerian orbits. In some cases, such studies
reveal complex problems with coplanar assumptions regarding the
system, particularly if such an assumption causes the system to be
unstable. A first-order analysis of the system is to determine the
proximity of the planetary orbits in relation to the Hill radii of the
planets. However, a thorough investigation via N-body numerical
simulations is needed to fully resolve such complex cases.

Our analysis of the HD~5319 and HD~7924 systems shows that they both
suffer from a fundamental instability based upon the precise published
orbital parameters, but long-term stable coplanar solutions may
still be found within the published one-sigma uncertainities for both
systems. Through an exhaustive suite of stability simulations that
varied the mutual inclinations of the outer planets, we have located
inclinations that satisfy system stability over a period of $10^7$
years. Depending upon the system, there may be several such islands of
stability in the parameter-space of inclination, keeping in mind that
lowering the inclination also increases the mass of the planet in
question. Our further analysis of the apsidal trajectory evolution for
stable mutual inclinations of the systems show that they can generally
be described as having circulating apsidal modes due to orbits of the
outer planets that are near MMR. The complexity involved in achieving
a full understanding of system stabilities is compounded by the
uncertainty in the Keplerian orbital elements. Further RV and
astrometric data for these and other similar systems will aid
enormously in resolving the close encounters evident in the system
architectures.

%%%%%%%%%%%%%%%%%%%%%%%%%%%%%%%%%%%%%%%%%%%%%%%%%%%%%%%%%%%%%%%%%%%%

\section*{Acknowledgements}

The author would like to thank Jonti Horner for useful discussions on
the stability simulations. Thanks are also due to the anonymous
referee, whose comments greatly improved the quality of the
paper. This research has made use of the following archives: the
Exoplanet Orbit Database and the Exoplanet Data Explorer at
exoplanets.org, the Habitable Zone Gallery at hzgallery.org, and the
NASA Exoplanet Archive, which is operated by the California Institute
of Technology, under contract with the National Aeronautics and Space
Administration under the Exoplanet Exploration Program. The results
reported herein benefited from collaborations and/or information
exchange within NASA's Nexus for Exoplanet System Science (NExSS)
research coordination network sponsored by NASA's Science Mission
Directorate.

%%%%%%%%%%%%%%%%%%%%%%%%%%%%%%%%%%%%%%%%%%%%%%%%%%%%%%%%%%%%%%%%%%%%


\begin{thebibliography}{}

\bibitem[Antoniadou \& Voyatzis(2014)]{ant14} Antoniadou, K.I.,
  Voyatzis, G. 2014, Ap\&SS, 349, 657
\bibitem[Bailer-Jones et al.(2013)]{bai13} Bailer-Jones, C.A.L.,
  Andrae, R., Arcay, B., et al. 2013, A\&A, 559, 74
\bibitem[Barnes \& Greenberg(2006a)]{bar06a} Barnes, R., Greenberg,
  R. 2006, ApJ, 647, L163
\bibitem[Barnes \& Greenberg(2006b)]{bar06b} Barnes, R., Greenberg,
  R. 2006, ApJ, 652, L53
\bibitem[Barnes \& Greenberg(2007)]{bar07} Barnes, R., Greenberg,
  R. 2007, ApJ, 6665, L67
\bibitem[Barnes et al.(2015)]{bar15} Barnes, R., Deitrick, R.,
  Greenberg, R., Quinn, T.R., Raymond, S.N. 2015, ApJ, 801, 101
\bibitem[Chambers(1999)]{cha99} Chambers, J.E. 1999, MNRAS, 304, 793
\bibitem[de Bruijne(2012)]{deb12} de Bruijne, J.H.J. 2012, Ap\&SS,
  341, 31
\bibitem[Duncan et al.(1998)]{dun98} Duncan, M.J., Levison, H.F., Lee,
  M.H. 1998, AJ, 116, 2067
\bibitem[Ford(2008)]{for08} Ford, E.B. 2008, AJ, 135, 1008
\bibitem[Fulton et al.(2015)]{ful15} Fulton, B.J., Weiss, L.M.,
  Sinukoff, E., et al. 2015, ApJ, 805, 175
\bibitem[Giguere et al.(2012)]{gig12} Giguere, M.J., Fischer, D.A.,
  Howard, A.W., et al. 2012, ApJ, 744, 4
\bibitem[Giguere et al.(2015)]{gig15} Giguere, M.J., Fischer, D.A.,
  Payne, M.J., et al. 2015, ApJ, 799, 89
\bibitem[Goldreich \& Schlichting(2014)]{gol14} Goldreich, P.,
  Schlichting, H.E. 2014, AJ, 147, 32
\bibitem[Howard et al.(2009)]{how09} Howard, A.W., Johnson, J.A.,
  Marcy, G.W., et al. 2009, ApJ, 696, 75
\bibitem[Kane et al.(2009)]{kan09} Kane, S.R., Mahadevan, S., von
  Braun, K., Laughlin, G., Ciardi, D.R. 2009, PASP, 121, 1386
\bibitem[Kane \& Gelino(2014)]{kan14b} Kane, S.R., Gelino, D.M. 2014,
  ApJ, 792, 111
\bibitem[Kane \& Raymond(2014)]{kan14a} Kane, S.R., Raymond,
  S.N. 2014, ApJ, 784, 104
\bibitem[Kane(2015)]{kan15} Kane, S.R. 2015, ApJ, 814, L9
\bibitem[Laughlin \& Chambers(2001)]{lau01} Laughlin, G., Chambers,
  J.E. 2001, ApJ, 551, L109
\bibitem[Lovis et al.(2011)]{lov11} Lovis, C., S\'egransan, D., Mayor,
  M., et al. 2011, A\&A, 528, 112
\bibitem[Marzari et al.(2006)]{mar06} Marzari, F., Scholl, H.,
  Tricarico, P. 2006, A\&A, 453, 341
\bibitem[Perryman et al.(2014)]{per14} Perryman, M., Hartman, J.,
  Bakos, G.A., Lindegren, L. 2014, ApJ, 797, 14
\bibitem[Petrovich et al.(2013)]{pet13} Petrovich, C., Malhotra, R.,
  Tremaine, S. 2013, ApJ, 770, 24
\bibitem[Rivera \& Lissauer(2001)]{riv01} Rivera, E.J., Lissauer,
  J.J. 2001, ApJ, 558, 392
\bibitem[Robinson et al.(2007)]{rob07} Robinson, S.E., Laughlin, G.,
  Vogt, S.S., et al. 2007, ApJ, 670, 1391
\bibitem[Smith \& Lissauer(2009)]{smi09} Smith, A.W., Lissauer,
  J.J. 2009, Icarus, 201, 381
\bibitem[Varadi(1999)]{var99} Varadi, F. 1999, AJ, 118, 2526
\bibitem[Wisdom \& Holman(1991)]{wis91} Wisdom, J., Holman, M. 1991,
  AJ, 102, 1528
\bibitem[Wisdom(2006)]{wis06} Wisdom, J. 2006, AJ, 131, 2294
\bibitem[Wright \& Howard(2009)]{wri09} Wright, J.T., Howard,
  A.W. 2009, ApJS, 182, 205

\end{thebibliography}
\end{document}